\documentstyle[epsfig]{./memsait}
\begin{opening}
\title{ON THE COUPLING OF ROTATION POWERED PULSARS TO PLERIONIC NEBULAE} 
\author{Jonathan Arons}
\institute{Department of Astronomy and Department of Physics\\
University of California, Berkeley}
\date{} 
\end{opening}

\begin{document}
\def \SAIT #1 #2 {{\em Mem.\ Soc.\ Astron.\ It.\/} {\bf #1}, #2}
\def \MESS #1 #2 {{\em The Messenger\/} {\bf #1}, #2}
\def \ASTRNACH #1 #2 {{\em Astron. Nach.\/} {\bf #1}, #2}
\def \AAP #1 #2 {{\em Astron. Astrophys.\/} {\bf #1}, #2}
\def \AAL #1 #2 {{\em Astron. Astrophys. Lett.\/} {\bf #1}, L#2}
\def \AAR #1 #2 {{\em Astron. Astrophys. Rev.\/} {\bf #1}, #2}
\def \AAS #1 #2 {{\em Astron. Astrophys. Suppl. Ser.\/} {\bf #1}, #2}
\def \AJ #1 #2 {{\em Astron. J.\/} {\bf #1}, #2}
\def \ANNREV #1 #2 {{\em Ann. Rev. Astron. Astrophys.\/} {\bf #1}, #2}
\def \APJ #1 #2 {{\em Astrophys. J.\/} {\bf #1}, #2}
\def \APJL #1 #2 {{\em Astrophys. J. Lett.\/} {\bf #1}, L#2}
\def \APJS #1 #2 {{\em Astrophys. J. Suppl.\/} {\bf #1}, #2}
\def \APSS #1 #2 {{\em Astrophys. Space Sci.\/} {\bf #1}, #2}
\def \ASR #1 #2 {{\em Adv. Space Res.\/} {\bf #1}, #2}
\def \BAIC #1 #2 {{\em Bull. Astron. Inst. Czechosl.\/} {\bf #1}, #2}
\def \JSQRT #1 #2 {{\em J. Quant. Spectrosc. Radiat. Transfer\/} {\bf #1}, #2}
\def \MN #1 #2 {{\em Mon. Not. R. Astr. Soc.\/} {\bf #1}, #2}
\def \MEM #1 #2 {{\em Mem. R. Astr. Soc.\/} {\bf #1}, #2}
\def \PLR #1 #2 {{\em Phys. Lett. Rev.\/} {\bf #1}, #2}
\def \PRL #1 #2 {{\em Phys. Rev. Lett.\/} {\bf #1}, #2}
\def \PASJ #1 #2 {{\em Publ. Astron. Soc. Japan\/} {\bf #1}, #2}
\def \PASP #1 #2 {{\em Publ. Astr. Soc. Pacific\/} {\bf #1}, #2}
\def \NAT #1 #2 {{\em Nature\/} {\bf #1}, #2}
%
\oddpagefooter{}{}{} 
\evenpagefooter{}{}{} 
\

\bigskip

\begin{abstract}
After discussion of observational constraints on the nature of the 
MHD wind coupling between the Crab Pulsar and the Crab Nebula, the theory
of transverse relativistic shock structure is reviewed and applied
to the interpretation of the wisps in the Nebula as the 
manifestation of the distributed wind termination shock structure, energetically
dominated by heavy ions, accelerated in the rotational equator of the
pulsar to energies comparable to the total voltage across the pulsar's
open field lines and carrying a current comparable to the Goldreich-
Julian current.  New results on the variability of the shock structure
are presented, which show that the gyrating ion bunches emit outwardly
traveling finite amplitude compressional waves, in agreement
with recent ground based observations.  The implications of the
theory for X-ray, $\gamma$-ray and high energy neutrino emission are 
briefly discussed, as are the problems of low magnetic energy density
in the upstream wind and the origin of the Nebular radio emission.
A brief discussion of other plerions leads to the conclusion that much
more detailed observations are needed before these systems can be 
modeled with the same sophistication as can be done for the Crab Nebula.
\end{abstract}

\section{Introduction}

Pulsars and soft gamma ray repeaters (SGRs), known to be neutron
stars with varying levels of confidence, are unresolved stellar sources.
Rotation powered pulsars lose most of their energy in an invisible
form; SGRs may also lose much of their energy in the same manner.  
Sometimes, this energy loss leaves its signature in the form of
unpulsed nonthermal emission from surrounding nebulae - the positional
identification of such nebulae with SGRs is the main argument for
identifying the SGR phenomenon with neutron stars (Kulkarni, this meeting).  
The nebular nonthermal
radiation is synchrotron and inverse Compton emission from relativistic 
particles and fields injected by the embedded pulsar.

The Crab Nebula is the most famous example of a plerion which forms a
box calorimeter around its pulsar. Other well known examples include
Vela-X (Bock {\it et al.} 1998) and 3C58 ({\it e.g.} 
Helfand {\it et al.} 1994), 
although in this latter nebula no pulsar has been
specifically identified. There are about 15 plerions 
identified from radio images, while recent advances in X-ray imaging
have found clear evidence for X-ray nebulae (presumably synchrotron
nebulae) around pulsars (Harrus {\it et al.} 1996).

When the pulsar's energy loss is relatively small and the pulsar's
space velocity is high, calorimetric nebulae enclosing the pulsar change their morphology and assume a cometary form. This change occurs
when the space velocity of the pulsar exceeds the 
expansion velocity the plerion would have if the pulsar were stationary with
respect to the interstellar medium. If the pulsar's true age equals
its characteristic age, then one expects to see cometary morphology or
other distortions when 
\begin{equation}
t_{char} \equiv \frac{P}{2 \dot{P}} > \frac{10^4}{v_{100}^{5/3} P^{2/3}} 
             \left( \frac{I_{45}}{n_{ISM}} \right)^{1/3} \; {\rm years} ;
\label{eq:comet-age}
\end{equation}
younger objects should be well embedded inside their nebulae.  Here
the rotation period $P$ is in seconds, $I_{45}$ is the moment of inertia
measured in units of $10^{45}$ cgs, $v_{100}$ is the pulsar's space velocity
in units of 100 km/s, and the interstellar particle number density is in
units of cm$^{-3}$. A subset of such ``plerionic'' nebulae forms within
binary systems, when the pulsar's outflow energy interacts with either the
companion star or with the mass loss from that star (Arons and Tavani 1993). 

The calorimetric nebulae surrounding young pulsars are of particular
interest, since understanding the physics of nebular excitation in these
systems can
yield reasonably unambiguous constraints on the physics of the
underlying pulsar, without the geometric confusion introduced by the
more complex flows around older pulsars. The Crab Nebula is still the only
system in which the quality and quantity of observational information
enables meaningful physical progress at a fundamental level.

\section{The Crab Nebula: The Interaction of PSR 0531+21 with the World}

This remnant of SN1054 emits nebular radio, IR, optical, X- and 
$\gamma$-rays ($\varepsilon < 100$ MeV), all of which is synchrotron
emission; higher energy photons probably are the Inverse Compton emission
of the same electrons (and positrons) that emit the
synchrotron radiation (de Jager and Harding 1992). The synchrotron 
lifetimes of the particles which
emit photons at near-IR and shorter wavelengths are less than the age of 
the nebula, thus requiring a continuous source of power, a requirement
fulfilled by the central pulsar, whose spin down luminosity
$\dot{E}_R \approx 5 \times 10^{38}$ ergs/s exceeds the total nebular
luminosity (primarily in X-rays and $\gamma$-rays) by about an order of
magnitude - the pulsar has $\sim$ 10\% efficiency in converting 
rotational energy loss into instantaneous
particle acceleration power in the Nebula. Furthermore, this power 
gets delivered to particles
whose maximum energy, as judged by the particle energies needed to
radiate 100 MeV synchrotron photons, is  $ \geq 10^{15.5}$ eV, comparable to 
the total voltage drop across the pulsar's open field lines.
The radiative lifetime of these 3 PeV electrons and positrons is quite
short, $\sim $ months; X-ray emitters lose energy in a few years. Thus the
high energy nebular emission provides a window into the pulsar's energetics
right now, demanding $10^{38.5} - 10^{39}$ electrons and positrons per
second from the pulsar, assuming the particles are accelerated only once.
The shrinkage of the nebular image with increasing photon energy shows that 
acceleration indeed does occur only once, and must be substantially
complete at radii no larger than the X-ray torus seen in ROSAT images,
{\it i.e.}, at projected radii less than 0.3 - 0.6 pc  
from the pulsar. The gamma ray source may well be smaller than the
X-ray torus, as is suggested by the hard X-ray image of
Pelling {\it et al.} (1987). If the acceleration process varies 
on time scales comparable to or longer
than the radiative lifetime of the $\gamma$-rays (for example), then the
gamma ray flux from the nebula will vary, since the short lived particles'
energy density and emissivity then follows the time variable acceleration
physics. EGRET data have suggested that some gamma ray variability does
occur on a time scale of months to years (de Jager {\it et al.} 1996).  

By contrast, the radio and far IR emitting particles have synchrotron
lifetimes greater than the nebular age. Therefore, their emission depends on
the history of the nebula, representing a convolution of the pulsar's
efficiency as a provider of accelerated particle energy, particle number and
magnetic field with the expansion history of the nonthermal bag. The expansion
history depends on the density and geometry of the external medium that confines
the relativistic particles and fields.  Averaging over the nebula's
history, the lower energy particles in the nebula show that the
pulsar has provided roughly $10^{40}$ synchrotron radiating particles/s. The 
discrepancy between this
average injection rate and the instantaneous injection rate to the X-ray torus
has not been resolved - the pulsar might have been a substantially more
active provider of relativistic particles in earlier epochs, or it might
provide an additional source of low energy particles today which is not
associated with the X-ray torus.
The nonthermal emission requires nonthermal nebular distributions of 
particles in energy space - ``power laws'', in the simplest representation 
of the data.

Images are essential to understanding the physics (radio, optical, UV,
X-Ray; hard X-ray and $\gamma$-ray if we could get them): for starters, 
the shrinkage of the nebular image with increasing photon
energy shows that the acceleration site is at or near the pulsar, not 
distributed throughout the nebula, at least for the higher energy particles.
Detailed imagery (radio, near IR, visual, soft X-ray) reveals the fine structure 
of the region where the pulsar rotational energy loss appears to be delivered 
to the nebula. As has been known since their discovery by Lampland (1921),
the NE-SW direction from the pulsar (the short axis of Nebula) shows ``wisps'', 
time variable (months to days) surface
brightness enhancements which are always present, between 5 and 30 arc seconds
from the pulsar (0.05 to 0.3 pc projected radius).  If one assumes the
nebula to be ``optically thick'' to the relativistic ram pressure
$\dot{E}_R /4 \pi r^2 c$ of the unseen outflow, balancing this ram pressure 
with the total nebular pressure ($p_{neb} \sim 10^{-8}$ dynes/cm$^2$) 
yields a termination radius for the unseen outflow right in the middle 
of the wisp region
(Rees and Gunn 1974), thus suggesting very strongly that the wisps are
an observational signature of the coupling between the pulsar and
the nebula. Recent ground based
optical studies have shown these variations appear to be waves
in brightness traveling outwards with speed $\leq 0.5c$ (Tanvir {\it et al}
1997), a result consistent with the preliminary results of the
ongoing HST imaging campaign (J. Hester and J. Graham, personal communications).
These variations are not correlated with timing glitches and other rotational
noise features of the pulsar, 
thus suggesting the variability is an intrinsic feature of the mechanism
which couples the unseen pulsar energy outflow to the nebula, rather than
being a passive consequence of variability in the pulsar's spindown.

The most efficient hypothesis is to assume the wisp region is the particle
acceleration zone, in addition to being the region where the pulsar outflow
energy becomes coupled to the nebular plasma. In this context 
``acceleration'' means the conversion
of the outflow energy, whatever it is, into the spectra of electrons
and positrons which are injected into the nebula.  As in the acceleration
of cosmic rays by supernova remnant shocks rather than by supernovae
themselves, such a hypothesis avoids the problem of the adiabatic losses
which plague mechanisms which rely on accelerating the observed particle
spectra within the pulsar's magnetosphere (e.g., Tademaru 1973). Granted that
the wisp variability time scale is months, that the $\gamma$-ray emitting
particles have radiative loss times on the same order, and that EGRET may
well have seen some gamma ray variability on month to year time scales, I am
much attracted by the hypothesis that the variable wisps are the direct
signature of the acceleration of particles to gamma ray emitting energies,
and thus are the site of the high energy electron and positron acceleration in the Crab Nebula.

\section{Wind Outflow From Crab Pulsar}

\begin{figure}
 
\centerline{\epsfig{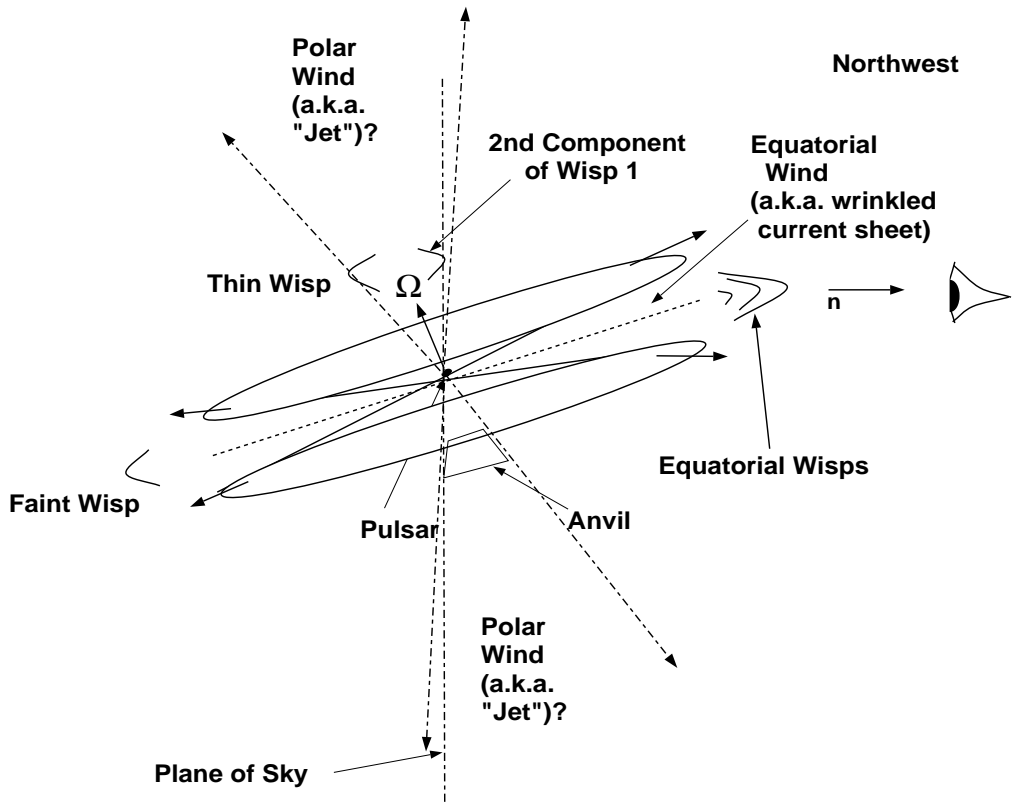}}
\caption {\label{fig:cartoon} The outflow geometry around the Crab Pulsar,
after Hester {\it et al.} 1995.}
\end{figure}

Figure \ref{fig:cartoon} 
shows a cartoon of the flow geometry near the Crab Pulsar. 
The X-ray morphology implies an outflowing ``disk wind'' in the equator, 
possibly associated with a corrugated equatorial return current sheet,
which terminates at or within the X-ray torus - in most interpretations,
the termination point identified depends on which feature the interpreter
adopts as a termination shock, implicitly assumed to be infinitesimally thin
on the scales of observational angular resolution.
The polar outflow
appears in the HST imagery as arcs both concave and convex toward the pulsar
(the thin wisp and the second, time stationary strand of wisp 1), interpreted
as possible shocks in a polar ``jet'' by Hester {\it et al} (1995). Within the
context of the fully MHD wind theories advanced by Kennel and Coroniti
(1984a,b), hoop stress in the toroidal magnetic field at high latitude
may compress the flow axially, contributing to the polar X-ray enhancements
 - the bolometric synchrotron surface brightness 
scales $\propto B^4$ (Woltjer 1958, Gallant and Arons 1994).  If so, the arc features
identified by Hester {\it et al.} might be interpretable as parts of
a polar termination shock's structure, in a manner similar to the theory of the
equatorial wind's termination described below. However, such speculations
are for future investigation, and will not be discussed further here.

It seems hard to avoid the conclusion that the outflowing energy feeding the 
X-ray torus has the character of a relativistic MHD wind.
The most widely accepted models of the electrodynamics of pulsars' polar caps
require an electric current along polar field lines with density
$J_\parallel \approx B/P$, which yields a total current 
$I_\parallel = 2 \mu \Omega_*^2 /c$ - here $\Omega_* = 2\pi /P$ and
$\mu$ is the magnetic moment. If the polar current couples to the wind/nebula
(an open circuit on light cylinder scales), the current induced $B$ makes a
considerable contribution to pulsar torque, a theoretical 
possibility supported  by the
fact that observed torques don't depend significantly on obliquity (Lyne and
Manchester 1988). In such open circuited models, the number of electric 
current carrying particles shot into the nebula per unit time is 
$\dot{N}_R = I_\parallel /Ze = 2 \mu \Omega_*^2 /Zec$, where $Z$ is the atomic 
number of the current carrying particles. Z=1 if the polar current is
electrons, as is the case when the obliquity between the magnetic moment
and the rotation axis is less than $90^o$, the geometry believed to be
appropriate for the Crab Pulsar ({\it e.g.} Romani 1996). 

Note that such models require another particle outflow from the neutron 
star to supply the required return current! For open circuited models, 
this outflow probably is
particles of the opposite charge sign extracted from an ``auroral ring''
around the polar flux tube, {\it ejected along the field lines that 
map into the rotational equator of the
PSR}, there to flow out in the (corrugated) equatorial current sheet. 
The return current in open circuited theory is still a cartoon - its
dynamics has yet to be explored. The results of shock theory applied
to the equatorial wind strongly suggest this return current 
in the Crab to be heavy ions, accelerated to the full
potential drop available (see below).

For the Crab, the number of electrons and positrons required to
feed the X-ray source tells us that at least in the equatorial wind,
the pulsar's loss rate in pairs is
\begin{equation}
\dot{N}_\pm \sim 10^{38.5-39} \; {\rm s}^{-1} \gg
   \dot{N}_R ({\rm Crab}) \simeq 10^{34.2} \; {\rm s}^{-1}.
\label{eq:pair-lossrate}
\end{equation}
It is well known that in pulsar flows, a total outflow rate of quasi-neutral
plasma in excess of $\dot{N}_R$ is a necessary condition for applicability
of relativistic MHD as the underlying theory (Arons 1979). 
Therefore, the simplest theory 
for the equatorial outflow is that the spindown energy loss is carried by a
relativistic MHD wind, with electrons and positrons as the
main constituents by number.  This wind might also contain a minority
population of heavy ions, which may or may not be energetically significant,
and may also carry part of the energy flow in the Poynting flux of the wound
up electromagnetic fields.

The rotational energy lost in nonradiative fields, pairs, heavy ions carried
by a MHD wind through the radiationless cavity at $r < 0.1$ pc from the pulsar
has its energy conservation described by
\begin{eqnarray}
\dot{E}_R  & = & r^2 \int d\Omega \left\{ \frac{c}{4\pi} {\bf E} \times {\bf B}
      + {\bf v}_{wind} \gamma_{wind} \left[(n_+ + n_- ) m_\pm c^2
      + n_i m_i c^2 \right] \right\} \cdot {\bf \hat{r}}  \nonumber \\
    & = & c\beta_{wind} \gamma_{wind} \dot{N}_i m_i c^2
       \left( 1 + \frac{m_\pm}{m_i} \frac{n_+ + n_-}{n_i}\right) (1 + \sigma),
\label{eq:energy-cons}
\end{eqnarray}
with the solid angle integration carried out over the sector of interest -
an equatorial sector with total latitudinal opening angle on the order of
$20^o$, if the wind feeding the observed X-ray torus has straight streamlines.
The parameters 
\newline 
\newline
$\bullet \; m_i\dot{N}_i$ = mass loss rate in ions 
\newline
$\bullet \; (n_+ + n_-)/n_i =  \dot{N}_\pm / \dot{N}_i$ 
  = ratio of pair number loss rate to ion number loss rate 
\newline
$\bullet \; \gamma_1$ = the bulk flow Lorentz factor  (or the velocity $v_1$) 
\newline
$\bullet \; \sigma$ = ratio of Poynting flux to kinetic energy flux in the wind
\newline
\newline
characterize the wind's properties.

MHD wind theory with $\sigma \ll 1$ gives a ``natural'' explanation 
of the Crab Nebula's dynamics\footnote{Recently Begelman (1998) 
proposed that MHD kink instability of toroidal magnetic fields allows
one to construct a wind fed model of the Crab with $\sigma \sim 1$ in the
wind, a conclusion which depends on the assumption, unstated in his
paper, that the kinked magnetic fields coagulate into patches whose 
filling factor in the nebula is small, and within which most of the field energy
annihilates.  Such coagulation is not a known consequence of the kink
instability (quite well studied in the low $\beta$ plasmas in fusion devices 
and the solar corona), and the virial theorem suggests such coagulation
to be unlikely. The observed uniformity of the radio spectral index
(Bietenholtz and Kronberg 1992) shows that the proposed annihilation must have little
radiative consequences for the radio emission in the Nebula, which is energetically
surprising.  If a kink instability does occur, a more likely 
consequence of kinking and reconnection 
would be to fill the Nebula with magnetic loops whose filling factor is
on the order of unity, in which case Rees and Gunn's original arguments for
low $\sigma$ in the wind are unaltered.  Nevertheless, the fine 
fibered structure observed by Scargle (1969) and by Hester {\it et al.} 
(1995) in the optical emission from the Nebula suggests some mechanism 
for complicating the magnetic structure on a fine scale is at work, for 
which Begelman's kink instability is a candidate.}(Rees and Gunn 1974, 
Kennel and Coroniti 1984):
\newline
\newline
$\bullet$ The deceleration of the post shock, low $\sigma$ pulsar wind 
($v_r \propto 1/r^2$) compresses the magnetic
field ($B \propto r$) until the magnetic energy reaches equipartition with the shocked 
relativistic plasma energy, whence deceleration and compression ceases
 - the model neatly explains equipartition as a dynamical effect;
\newline
$\bullet$ The fit to the observed expansion velocity requires  
   $\sigma << 1$ (caveat - Kennel and Coroniti neglected possible inertial 
loading by the filaments); 
\newline
$\bullet$ If the wind termination shock is assumed to create 
power law distributions of pairs, with upper and lower cutoffs determined
by the jump conditions and the particle spectral index at the shock constrained
by the final fit to the data, the global optical, X- and gamma-ray spectrum of the Crab can be reproduced by the model, once synchrotron cooling in the flow
is properly incorporated (Kennel and Coroniti 1984b).

But, radio emission from the Nebula is entirely left out! The inferred 
particle injection rate ({\it now}, since rapid synchrotron losses
make the X-rays a calorimeter for the current injection rate) is an order of magnitude smaller than the rate of injection of radio emitting particles, averaged over the life of the Nebula. Thus, the main stored component of the
relativistic energy in the Nebula was neglected. The non-spherical morphology
was not quantitatively addressed. Conceivably, the problem of radio  
emitting particle injection is related to  
the strong latitudinal asymmetries revealed by HST and other imaging.

\section{Physics of Relativistic Shock Structure/Dissipation/Acceleration}

As remarked above, dynamic pressure balance puts the termination of the
pulsar's unseen outflow at $r_s \sim 0.1 - 0.2$ pc (Rees and Gunn 1974). 
Shock dissipation is the most likely wind termination mechanism in the MHD theory. 
A shock in a MHD outflow from the Crab pulsar must be transverse: 
$\angle ({\bf B}, {\bf v}) \approx \pi/2 - 10^{-9} \Rightarrow$
diffusive Fermi acceleration has little relevance to the conversion of flow
energy into power law distributions of particles downstream. 
The phenomenological shock model with the shock regarded as  
infinitesimally thin and located at $r_s \approx r_{\rm wisp}$ 
simply requires the 
conversion of $\sim 10$\% of the flow energy into the spectra of 
pairs with distribution functions immediately downstream of the shock
$N_{\rm injected} (\gamma) \propto \gamma^{-s}, \; s \approx 2.2$; it
doesn't explain how the system achieves this efficiency.

My contributions to the subject come partly under this heading.
The results, based on linear instability theory, particle-in-cell
simulations and a modicum of quasi-linear theory, were mostly published some
time ago (Alsop and Arons 1988, Langdon {\it et al.} 1988,
Hoshino and Arons 1991, Gallant {\it et al.} 1992, 
Hoshino {\it et al.} 1992). Those interested in the detailed support of most
of the assertions made in this section should consult the papers referenced.

Imagine what happens when the incoming flow ``collides'' with the
magnetic step formed by the shock. The particles {\it all} reflect 
coherently, and start gyrating coherently within the shock front, now
considered more realistically as a transition region of finite
thickness in the flow. The coherently gyrating particles radiate
\underline{collective} cyclotron and synchrotron waves with fundamental 
frequencies $\omega_1 = eB/mc\gamma_1$, as well as
large amounts of power at the harmonics $\omega_l = l \omega_1$, 
including the high harmonics $l \gg 1$. The basic mechanism is azimuthal
bunching of the ring distributions in momentum space set up by the coherent
reflection from the magnetic step, whose shape is self-consistently
maintained by the particle rings in momentum space.
The pairs radiate extraordinary modes, with 
$\omega_1 \geq (eB_1/m_\pm c \gamma_1) \sigma^{-1/2} \sim 10^{-3} \; s^{-1}$.
The numerical value assumes
the upstream magnetic field to be $B_1 \sim 10^{-4.5}$ Gauss, 
the upstream flow Lorentz factor to be $\gamma_1 \sim 10^6$ and 
$\sigma \sim 10^{-2.5}$, all values taken from the Kennel and Coroniti model
or from the Gallant and Arons (1994) model described below. Coherent
gyration of heavy ions as they encounter the magnetic step
excites transverse magnetosonic waves which propagate with properties
mainly determined by the pairs (if the pairs are numerically in the majority,
as turns out to be the case), with frequencies
$\omega \geq ZeB_1/m_i c \gamma_1 \sim 10^{-6.5} \; {\rm s}^{-1}$,
a gyration time of months\footnote{In the actual application to the Crab,
the ions gyrate in a $B$ field already compressed by a factor of two to three
above its upstream value by the preliminary shock in the pairs, which
increases the gyration frequency by the same factor and yields an ion
gyration time of 1-2 months.}.

Cyclotron reabsorption of the extraordinary modes at the shock's leading edge
thermalizes the pairs to a relativistic Maxwellian distribution
with downstream temperature $T_\pm \approx \gamma_1 m_\pm c^2$ - the
mean free path for extraordinary mode emission and absorption in the 
pair plasma is much smaller than the flow scale length, leading to
the establishment of local thermodynamic equilibrium for essentially the
same reason that the emission and absorption of virtual photons (Coulomb
collisions) establishes LTE in a collisional nonrelativistic
plasma - in the relativistic case, the pairs and their waves form a local
{\it hohlraum}. The relativistic cyclotron instability in the ions 
simply serves to
establish a level of electromagnetic fluctuations (corresponding to real
photons in this case) at the thermal level far faster than would occur
if two body encounters were the only means of creating the fluctuating
electromagnetic field.  The simulations 
show that when pair thermalization is complete, the radiation level also
corresponds to LTE (in the Rayleigh-Jeans limit, as is expected for these
classical investigations).

From the perspective of the pairs, the magnetosonic waves emitted by the
more slowly developing relativistic ion cyclotron instability 
are an external source of energy, which can upset their thermal 
equilibrium.  The simulations show that the magnetosonic waves
have a nonthermal spectrum, basically corresponding to $1/f$ noise.
These waves are preferentially cyclotron absorbed by the more mobile
pairs, first at ion harmonics $l \sim m_i /Z m_\pm$, then, as the pairs gain
energy and detune from the high harmonics, from waves in the power law 
spectrum with successively lower frequencies, until acceleration 
stops for pairs whose energy equals that of an upstream ion,
for which the cyclotron frequency equals that of the ions that drive the
acceleration.

Indeed, a simple application of quasi-linear theory to this process (Arons,
unpublished) shows that the acceleration rate of an electron or a positron
in a spectrum of linearly polarized magnetosonic waves 
\begin{equation}
U_k =  r_{Li} \frac{(\delta B)^2}{4\pi} (kr_{Li})^{-2},
\end{equation}
the spectrum exhibited by the simulations, is
\begin{equation}
\frac{\dot{\gamma}}{\gamma} = \frac{0.017}{n_0^3} 
      \left(\frac{\delta B}{B} \right)^2 \Omega_{Li}.
\end{equation}
Here $k$ is the wavenumber, $U_k dk$ is the wave energy density in 
the interval ($k, k+dk$), $r_{Li} = c/\Omega_{Li}$ is an ion's Larmor
radius and $\Omega_{Li}$ its relativistic Larmor frequency, and
\begin{equation}
n_0 = \left(\frac{c^2 + v_A^2}{c_s^2 + v_A^2} \right)^{1/2},
\end{equation}
is the index of refraction of low frequency, small amplitude magnetosonic waves in
the pairs, with $c_s$ the relativstic sound speed in the pairs and $v_A$
the \underline{relativistic} Alfven speed in the relativistically
hot pairs. Typically $n_0 \sim \sqrt{2} - \sqrt{3}$. 
Formal applicability of quasi-linear theory requires
$\delta B /B \ll 1$, while the simulations show that $\delta B /B >2$
for parameters of interest.  Nevertheless, the simulations show that
the acceleration rate is indeed Fermi-like, with 
$\dot{\gamma} / \gamma \sim \Omega_{Li}$ - 
quasi-linear theory yields the correct scaling of the rate for the 
resonant process even when the fluctuation amplitudes are large, 
although its estimate of the numerical value is less reliable.

Subjected to the nonthermal heating of this resonant absorption process,
with losses being simply outflow of pairs from the region where the ions
lose their energy to the pairs, 
the relativistic Maxwellian pairs downstream from the pair shock develop
a nonthermal distribution
\begin{equation}
N_\pm (\gamma) \propto \gamma^{-2}, \; 
    \gamma_1 < \gamma < (m_i/Zm_\pm) \gamma_1.
\end{equation}
For lower energies, the particle spectrum remains that of a relativistic
Maxwellian.  The efficiency of energy transfer from the ions to the
pairs is
\begin{equation}
\varepsilon_a = \frac{{\rm nonthermal \; pair \; energy}}
    {{\rm total \; upstream \; flow \; energy}} \approx 10 - 20 \% ,
\end{equation}
a result known solely from simulation.  These acceleration
results obtain when the
ions provide the largest component of the upstream flow energy,
and are remarkably like those inferred from
application of ideal MHD shock theory to the Crab Nebula, with power law 
populations of pairs \underline{assumed} in the post shock flow.

\section{Wisps as Internal Shock Structure}

The physics of these shocks implies a model for the coupling of the
equatorial wind outflow from the Crab pulsar. Suppose the equatorial flow to
be composed of $e^\pm$ pairs, heavy ions and wound up magnetic fields 
in an unknown mixture,
all flowing out from the pulsar at super-Alfvenic speed. The shock
thermalizes the pairs to a relativistic Maxwellian distribution within the leading edge of the shock structure - this thermalization region has
radial thickness $\sim r_{L\pm} \sim 10 $ AU, unobservably thin to all but
VLBI radio observations. The ions have much larger gyration radius in the
compressed magnetic field supported by the shock heated pairs
($r_{Li} \sim  0.3 \; A/Z$ pc).  Relativistic cyclotron
instability of the gyrating ions generates large amplitude, long wavelength 
($\lambda \sim 0.2 /l$ pc, $l \geq 1$),
{\it compressional}, linearly polarized magnetosonic waves in the heated pairs.  
Cyclotron absorption of these waves causes gradual nonthermal 
acceleration of the pairs over a length $\sim$ several ion Larmor 
radii\footnote{This length is the distance required to bring the highest 
energy pairs in the spectrum to their maximum energy $\sim \gamma_1 m_i c^2$,
at which energy they radiate 100 MeV gamma rays. The nonthermal spectrum of
infrared and optically emitting particles is established within a distance
$ \sim 10 r_{L\pm} \approx 100$ AU downstream of the leading edge shock in the
pairs. Therefore, one needs better than 10 mas angular 
resolution to allow detection of the initially thermal O/IR synchrotron spectrum.}.
The ions follow a coherent orbit for a couple of gyration cycles, becoming 
progressively more disorganized as the instability broadens their momentum
distribution and drains their energy into magnetosonic waves.  However,
in the first few cycles of coherent gyrational flow, the turning points in 
the ion orbits are coherently spaced with separations $\sim r_{Li}$. Since
the radial outflow (and inflow) momentum of the gyrating ions must be deposited
in the magnetized pairs as the ions gyrate, the turning points 
correspond to compressions in the magnetic field and pair plasma. 
Such compressions correspond to surface brightness enhancements 
separated in radius by the ion Larmor radius scale, with bolometric synchrotron emissivity
$\propto B^4$. The compressions also couple the ion momentum to the
propagating magnetosonic waves of the pair plasma, since the cyclotron
instability makes the ion reflection process time dependent in the ion
drift frame. Thus the compressions created by the reflected ions should travel
outwards in the pair flow frame with the magnetosonic speed of the nonlinear waves.

The properties of such compressions have a not unreasonable similarity to
the observed properties of the wisps, suggesting that the wisps are the
observable manifestation of the {\it internal structure} of an (energetically)
ion dominated shock terminating the equatorial wind from the Crab pulsar.  
If this hypothesis is correct, the whole shock structure
is spread across the sky, turning the Crab Nebula into a laboratory for
the relativistic shock physics believed to be central to a wide variety
of high energy astrophysical systems.

\begin{figure}
\centerline{\epsfig{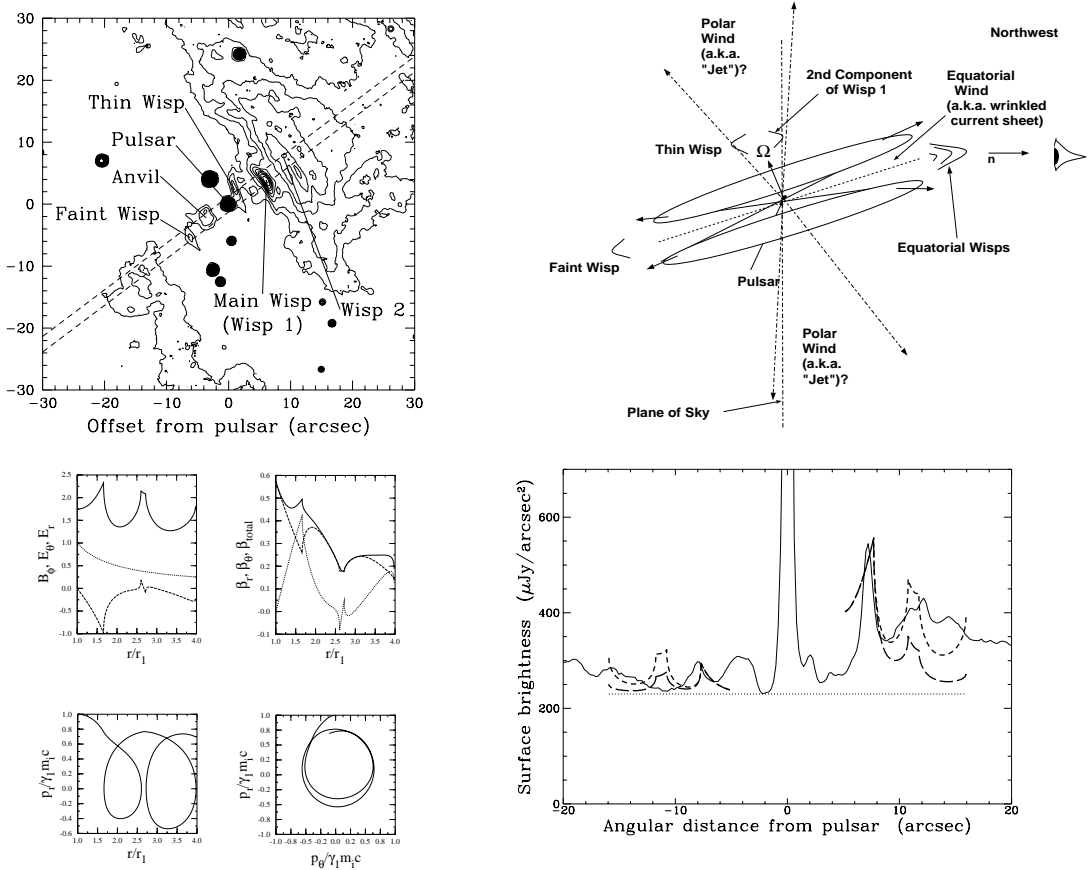}}
\caption[h]{\label{fig:ga-model} The Gallant and Arons (1994) model.
Upper left panel: The van den Bergh and
Pritchet 1989 image of the Crab's central regions.  The surface brightness
in the strip between the dashed lines was used to provide the data with which the
model is compared. Upper right panel: The geometry of the region around
the pulsar, as in Figure \ref{fig:cartoon}. Lower left panel: The 
theoretical structure of the  model - upper left: B field (top curve), transverse E
field (middle curve), radial E field (bottom curve); upper right: 
flow velocity of the pairs 
(top = total, middle = radial, bottom = transverse current velocity); 
lower left: radial ion momentum as a function of position; lower right:
ion phase space, showing the laminar ring orbits.
Lower right panel: synthetic surface brightness compared to the observations: dashed 
curve with large central peak - the I-band data from van den Bergh and Pritchett 
(1989); remaining dashed curves - the models, assuming thermal synchrotron emission from
the pairs, for two different assumptions about the rate of pitch angle broadening
of the anisotropic pair distributions. Note the failure of the model to account
for the thin wisp, for the bifurcated structure of wisp 1 and for the anvil, 
failures which are all a consequence of the fact that these extra features 
are due to superposition of features in the polar flow on the images of the
equatorial structures when projected onto the plane of the sky.}
\end{figure}

Gallant and Arons (1994) decided to test the configuration space
aspects of the model outlined above by constructing a quantitative 
{\it steady} flow 
theory, assuming the flow to be confined to a sector of a sphere within
latitude $\sim \pm 10^o$ of the pulsar's rotational equator.  
The ion flow was modeled as a {\it laminar} stream of particles 
with no momentum dispersion, gyrating
in the magnetic field embedded in a shock heated, Maxwellian pair fluid
whose flow was modeled as adiabatic - no attempt was made to model either
the nonthermal particle acceleration or the variability observed in the
simulations, and by construction the model creates compressions which are stationary
in space - the pairs flow through these standing waves.  This model and 
its quantitative results, when applied to the I-band snapshot of 
van den Bergh and Pritchet (1988), are illustrated in Figure 
\ref{fig:ga-model}.  

Among the model's highlights are a simple explanation of the NW-SE
brightness asymmetry of the wisps as being due to the Doppler boost in the
mildly relativistic pair flow in the ion gyration region. The parameters
inferred from the best fit of the model to the main wisps 1 and 2 
in the NW and the faint wisp in the SE suggest this steady, reflected 
ion flow model has not unreasonable correspondence to van den Bergh 
and Pritchett's observations of wisp separation, brightness and
shape, which overdetermine the model:
\newline
\newline
\begin{tabular}{l}
$\sigma \approx 3 \times 10^{-3}, $ \\
$\gamma_1 \approx 4 \times 10^6 \approx 0.3 Ze \Phi_{open} /m_i c^2, $ \\
$B_1 \approx 3 \times 10^{-5} \; {\rm Gauss},$ \\
$\dot{N}_\pm \approx 10^{38} \; {\rm s}^{-1}, $ \\
$m_i \dot{N}_i \approx 2 m_\pm \dot{N}_\pm \approx 10^{-15} \; {\rm M_\odot /yr}
         \approx 50,000 \; {\rm metric \; tons /s}$,  \\
$Z \dot{N}_i \approx 3 \times 10^{34} \; {\rm s}^{-1} 
         \approx {\rm Goldreich-Julian \; return \; current}$.\\
\end{tabular}
\newline
Here $\Phi_{open} = \sqrt{\dot{E}_R /c} = 4 \times 10^{16}$ Volts is the
total electric potential drop across the open magnetosphere.  The fit to the 
data also fixes the ion Larmor radius and the tip angle of the equatorial outflow
to the line of sight: $r_{Li} \approx 0.15$ pc and $\angle$(LOS, equatorial
wind) $\approx 35^o$. The number of pairs flowing out in the equatorial
wind is close to, but somewhat less than what was inferred in the
Kennel and Coroniti model as the particle supply needed for the X-ray
source, and the ion current is in good
agreement with what one would expect if the equatorial wind carries
the return current.

While the favorable comparison of the Gallant and Arons model 
to a single, relatively
low resolution optical snapshot of the Crab gives some credence to the basic
idea that the wind is energetically dominated by very high rigidity ions, the neglect of time dependence is a serious flaw in modeling the equatorial outflow.  
Variability of the equatorial wisps has been known since Lampland's 
original discovery, seen most spectacularly in the high resolution
``movie'' created from the HST campaign now in progress 
and in the ground based optical observations reported by Tanvir {\it et al.}
(1997).  Both sets of data show that the wisps are outwardly
propagating structures, behaving like spherical or cylindrical waves
which lose coherence over several 
tenths of a parsec as they propagate away from the pulsar.

The kinetic simulations of ion dominated relativistic shocks in plane parallel 
geometry published by Hoshino {\it et al.} (1992) (see also Hoshino 1998)
clearly show the shock structure to be time dependent, with a large amount of 
short wavelength power in the magnetic field.  The basic 
relativistic cyclotron instability of the shock structure implies
variability of brightness enhancements on the ion gyration time scale, with
faster variability imposed on the basic structure by the higher harmonics. Such
variability would be uncorrelated with pulsar timing variations, as seems
to be the case in all the observations of wisp variability. However, the
original simulations did not give a clear answer to whether the variability
is in the form of oscillations of the shock structure around a
mean position, or in the form of ``radiation'' of finite amplitude
magnetosonic waves into the surrounding nebula.

An investigation of the time dependent theory (Spitkovsky and Arons, 
in prep) shows that the relativistic cyclotron instability of the
ion ring formed immediately downstream of the thin shock in the
pairs {\it does} launch outwardly running waves in the magnetic field, density 
and temperature of the pairs (contrary to the criticism of this model advanced 
by Tanvir {\it et al.} 1997, who assumed that shock variability corresponds to
shock oscillation around a mean position), with fundamental
period comparable to the ion Larmor time 
$t_{Li} = 2 \pi c/\Omega_{ci2} = m_i c \gamma_1 /Ze (3B_2) \approx 1.5$ months.

\begin{figure}
\centerline{\epsfig{file=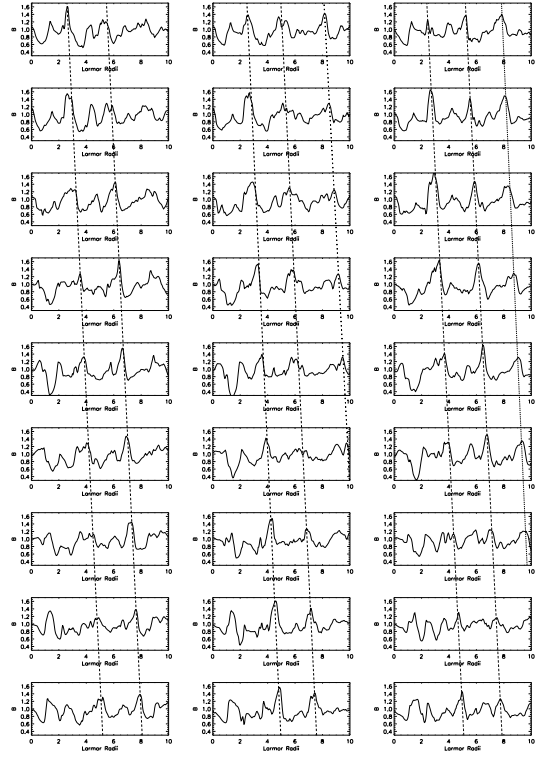,height=4in}}
\caption[h]{Snapshots of the magnetic field as a function of position,
from a plane-parallel hybrid simulation of the relativistic cyclotron
instability of an ion beam injected into a relativistically hot $e^\pm$
magnetized plasma itself moving with speed $v_x = c/3$, 
with the $B$ field transverse to the flow. The snapshots are read starting
from the top of the left column, then continuing to the second and 
third columns. The time between each snapshot
is $t_{Li} / 18$. The sloping lines connect identifiable features in the
wave profile, which themselves vary in the frame traveling with the
wave group velocity. The slopes of these lines provide the wave speed in
the simulation frame. \label{fig:hybrid}}
\end{figure}

These calculations use a ``hybrid'' approach to modeling the shock
structure. Particle-in-cell ions are injected into hot (Maxwellian) pairs, 
modeled as an ideal, relativistic MHD fluid with magnetic field transverse 
to the flow.  The pair fluid is still modeled as adiabatic, with 
cyclotron absorption of the  ion waves and  nonthermal pair acceleration
neglected. Some illustrative results from a plane parallel
simulation are shown in Figure \ref{fig:hybrid}, which shows the compressions
in the $B$ field that form as the ion ring breaks down into time dependent bunches 
in gyrophase.  The compressions in the pairs and $B$ field observed in the
code travel downstream with speed 
$v_{\rm wave} \approx 0.8c$; theoretically, \underline{small} amplitude waves should
travel with speed\footnote{The adiabatic index of the pair fluid used in
the simulation was $\Gamma = 3/2$, as would be the case if the pairs were
heated only in the plane orthogonal to ${\bf B}$.}
$$v_{\rm wave} = \frac{v_{pairs} + v_{ms}}{1 + \frac{v_{pairs} v_{ms}}{c^2}}
   = \frac{\frac{c}{3} + \frac{c}{\sqrt{2}}}{1 + \frac{\frac{c}{2}
		\frac{c}{\sqrt{2}}}{c^2}} = 0.84c.
$$

Here $v_{ms}$ is the wave group speed in the proper frame of the pair fluid.
The numerical value is peculiar to the plane parallel geometry used in
this particular calculation (mainly to test the code). These
preliminary results suggest that the interpretation of the wisp region
as the equatorial shock structure remains viable.

\section{Speculations and Conclusions}

These dynamic models of distributed shock structure continue to show
promise in the interpretation of the energy transfer between the Crab
pulsar and its nebula.  A complete theory requires addressing a variety
of other issues, which one can only do at the order of magnitude level
at present.

\subsection{X-rays and Gamma Rays}

The plane parallel shock structure and acceleration calculations, and 
quasi-linear theory applied to those simulations,  yield 
the maximum energy of the power law formed in the downstream pairs
to be $E_{\pm , max} \approx (E_{ion})_{upstream} \approx  0.3 Ze\Phi_{open}
\approx 10^{16}$ eV. The same calculations yield an
acceleration time to extend the power law  to the maximum energy
to be $t_a ( E_{\pm , max}) \approx \Omega_{Li2}^{-1} \approx 1.5$ months.
The resulting \underline{synchrotron} spectrum rolls off above the photon
energy
$\varepsilon_2 \approx 0.3 (\hbar e B_2 /m_\pm c) 
    (m_i /Z m_\pm)^2 \gamma_1^2 \approx 20$
Mev, using the Gallant and Arons parameters, a value quite close to the
rolloff energy of the variable $\gamma$ ray component of the spectrum
reported by de Jager {\it et al.} (1996). The synchrotron loss time at 
these highest energies is $t_s (E_{\pm,max}) \sim 4$ months, and the size of the
synchrotron source in the 10-100 MeV region then should be
$R_s \leq v_{pairs}t_s \simeq 2 \times 10^{17} \; {\rm cm}$.  Furthermore,
the spatially progressive nature of the particle acceleration 
suggests that the size of
the source at the highest energies will stop shrinking on scales smaller 
than about $10''$, and that interior to several arc seconds from
the pulsar, the higher energy emission should show a ``hole'' in the surface
brightness, with the hole size {\it decreasing} with decreasing photon energy,
until the leading edge shock in the pairs is reached.

Because the shock structure is unsteady, the particle acceleration also
varies.  When the synchrotron loss time greatly exceeds the variability
time scale of the accelerator introduced by the large amplitude magnetosonic 
waves, the radiation
physics averages over the variable particle acceleration. At the highest
photon energies, however, the synchrotron loss time is comparable to
the fundamental magnetosonic wave time scale, suggesting that the 
50-100 MeV \underline{synchrotron} source
varies with fractional luminosity changes
$$
\frac{\delta L_{\gamma, synch}}{L_{\gamma , synch}} \approx
     \frac{t_a ( E_{\pm , max})}{t_s( E_{\pm , max})} \approx 0.4.
$$
At these photon energies, the inverse Compton source overlaps the
synchrotron source in energy space. Since the inverse Compton source must
have size comparable to the optical Nebula (de Jager and Harding 1992), 
there will not be
variations of the inverse Compton radiation on these short time scales,
with the result that the \underline{total} 50-100 MeV emission will vary less
than the synchrotron component alone. Nevertheless, these predicted variations
on the several month time scale should be marginally observable in the
EGRET data.  Also, a substantial, improvement of angular resolution in hard
X-rays will soon allow new probes of the Crab's inner workings, when the
HESSI satellite with its $2''$ imaging at several hundred keV is launched.

\subsection{The Radio Nebula}

Optical, X- and $\gamma$-ray emission diagnoses the coupling physics ``today''.
The radio emission measures the integral of the pulsar's input
over history - most of
the stored relativistic energy is in $B$ fields and radio emitting
particles ($\sim 10^{50}$ ergs). Averaged over the whole Nebula, the radio
emitting spectrum has the form 
$N_\pm(\gamma) \propto \gamma^{-1.5}, \; 10^{2.5} < \gamma < 10^4$; indeed,
detailed spectral index maps (Bietenholz and Kronberg 1992) show this
particle distribution to be remarkably homogeneous.

The wind termination shock models constructed to explain the
\underline{equatorial} wind (Kennel and
Coroniti 1984a,b, Gallant and Arons 1994) don't yield 
$N(\gamma) \propto \gamma^{-s}, \; s \sim 1.5$ at energies
small compared to $10^6 m_\pm c^2$. Also, the pair injection rate
inferred in the equatorial wind is between a few  and 10 per cent of the average
injection rate needed to explain the radio Nebula.

This discrepancy poses the following conundrum. Spectral continuity 
suggests one mechanism accelerates the synchrotron emitting particles, yet the shock
jump conditions applied to the equatorial wind clearly show the 
particle spectrum must be ``cut off'' below $10^6 m_\pm c^2$ (Kennel
and Coroniti 1984b); in the kinetic theory of the shock acceleration physics,
the particle spectrum remains Maxwellian below $\gamma_1 m_\pm c^2$
(Hoshino {\it et al.} 1992), which yields the same low frequency emissivity
as does a sharply cut off distribution, $f_\nu \propto \nu^{1/3}$, not the
observed $f_\nu \propto \nu^{-0.25}$.  It is possible that additional
acceleration physics within the shock structure beyond cyclotron resonant
absorption of the magnetosonic waves can lead to a nonthermal low energy
spectrum like that observed - magnetic pumping is an interesting candidate.
But, modifications of the acceleration physics will not alleviate the
discrepancy between the rate of pair injection into the equatorial torus and
the average injection rate of radio emitting electrons.

The discovery of the ``polar jet'' (Hester {\it et al} 1995)
suggests that wind outflow at latitudes $|\lambda | > 10^o$, which could
fill most of the solid angle around the pulsar, might provide the source
of the larger number of particles feeding the radio source.  If
the acceleration physics is the same, perhaps spectral continuity is not a 
surprise. However, one would be quite surprised to find the wind
outside of the magnetic equator to contain an energetically dominant
component of heavy ions, for the electrodynamical reasons described earlier.
Since the energetic dominance of the ions is essential to the cyclotron
resonance acceleration explanation of the nonthermal pairs in the
equatorial flow, invoking the same acceleration physics at higher latitudes
doesn't look to me to be the right theoretical path.  One is left with the
possibility that the higher latitude acceleration physics is different,
or that radio electrons are accelerated in the outer Nebula (for example,
by Fermi I acceleration at the shock just outside the synchrotron Nebula,
whose presence was inferred by Sankrit and Hester  1996), and that spectral
continuity is just a coincidence.  Observations of the spectra of other
plerions  with sufficient spectral coverage (radio, IR, optical, X-rays)
would greatly help in directing the course of theory - if most plerions
show substantial spectral discontinuities between radio and higher
frequencies, then the Crab's spectral continuity clearly would
be coincidental. But if the spectra all show continuity similar to the
Crab's, then a solution must be sought in terms of a more unified
injection/acceleration scheme than is implied by current theory. 

\subsection{High Energy Neutrinos from the Crab Nebula}

In principle, observation of high energy neutrinos from the
Crab would be a direct test of the ion doped wind model (and of
other ideas concerning ultra high energy ion acceleration
by the pulsar, such as the outer gap construction of Bednarek and Protheroe 
1997). The following numerical example illustrates the possibilities.

Suppose the pulsar injects heavy ions at the Goldreich and Julian rate,
$\dot{N}_i = 10^{34.5}/Z \; {\rm s}^{-1}$, each with energy
$\gamma_i m_i c^2 = \eta Ze \Phi_{open} = 4 \times 10^{16} Z\eta$ eV.
The ions' Larmor radii in the nebula are 
$r_{Li} = \gamma_i m_i c^2 /ZeB_{neb} = \eta \Phi_{open}/B_{neb} 
\approx 0.15 \eta (10^{-4} \; {\rm Gauss} /B_{neb})$ pc. The ions drift
out of the Nebula with drift velocity across $B$
$\sim c r_{Li} /R_{neb} \sim 0.1 c$, a speed about 10 times the hydrodynamic
expansion velocity of the nebula. Then the number of ions contained 
in the Nebula is 
$N_i = \dot{N}_i (R_{neb} /c)(R_{neb} /r_{Li}) \sim 6 \times 10^{43} Z^{-1}.$
With a cross section for $\pi^\pm$ production on the order of 10 mb, and
with $\sim 1 \; M_\odot$ of thermal material within the nebula (contained
in the emission line filaments, corresponding to an average gas 
density of a few protons/cc), the high energy neutrino luminosity
of the Crab Nebula should be at least $10^{29}$ high energy neutrinos/s.
This is a lower limit, since the heavy ions continue to interact with
the invisible gas confining the visible synchrotron nebula as they
to wander outwards through the unknown magnetic field in the inertially
confining material. At 2 kpc distance, the neutrino flux from the Crab
should be about $5 \times 10^{-15}$ neutrinos/cm$^2$-s, substantially
above background at the $\sim 10^{16}$ ev energy suggested by this elementary
monoenergetic model. At these high energies, the count
rate expected in the proposed high energy neutrino observatories is
rather low. However, the use of the Crab Nebula as an ion calorimeter 
through the dynamics of the wisps does
not tell us anything about the downstream spectrum of injected ions, and theory 
of ion acceleration at the pulsar is
too primitive to be of much help.  If the ions have a sufficiently steep
injection spectrum, the production rate of lower energy neutrinos might
be much higher.

\subsection{Why is $\sigma \ll 1$?}

One of the most surprising conclusions of the Rees and Gunn (1974) model
of the Crab is that $\sigma \ll 1$ in the wind upstream of the termination
shock, a result which has persisted in all of this model's descendents.
Since the pulsar's magnetosphere is magnetically dominated ($\sigma \gg 1$),
and $\sigma$ is conserved in simple ideal MHD wind models, the fate of
the pulsar's magnetic energy has attracted quite a bit of theoretical attention.
Suggestions which appear to have some possibility of success 
include dissipation of the magnetic field in a striped 
MHD wind by tearing modes (Coroniti 1990); dissipation of
the magnetic field in a striped MHD wind because of insufficient
current carriers to support the stripes (Michel 1994); 
and conversion of MHD flow to dissipative ``vacuum'' waves in the wind zone
(Melatos and Melrose 1996). All of these ideas depend upon most of
the magnetic field in the wind having a wave-like structure, either
as standing oscillations in the fluid frame (Coroniti, Michel), or as
large amplitude waves which propagate with respect to the plasma
in the fluid frame, with structure dominated by displacement
current even though the waves are subluminous (see also Melatos' paper
in these proceedings). If any of these thoughts is on the right track,
the low $\sigma $ problem requires giving up the applied mathematical
pleasures of studying the aligned rotator and axisymmetric ideal MHD winds -
indeed, one has to give up ideal MHD!  None of the proposed ideas, however,
has been developed to the point of usefully confronting theory with
the elaborate HST pictures or other high resolution observations.

\subsection{Other Models of the Wisps and the Pulsar-Nebula Coupling}

I would be remiss if I neglected discussing some of the other
ideas around for the interpretation of the Crab's wisps and what they have
to tell us about the pulsar-nebula connection.  Woltjer (1958) is the first
to suggest the idea that the wisps might be damped magnetosonic waves, perhaps
driven by Baade's star, the then mysterious object suspected of having 
something to do with the energization of the Crab Nebula. Scargle (1969) and 
Barnes and Scargle (1973) rediscovered this idea, and proposed the
wisps to be magnetosonic waves in a relativistic electron-heavy ion plasma, 
launched by upstream variations of the vacuum magnetic dipole
radiation then thought to carry the pulsar's spin down energy. They
attributed the time variability of the magnetic dipole radiation
timing glitches, not to
intrinsic instability in the termination of the strong wave, while
the nebular relativistic electron spectrum was attributed to  
Landau damping (``Barnes damping'') of the waves in the Nebular plasma. 
Unfortunately, the wisp variability does not correlate with glitches,
and Barnes damping does not lead to the observed particle spectrum.
Nevertheless, the ideas expressed by these authors have a clear
relationship to the model I outlined above.

Hester  {\it et al.} (1995, and personal communications), has expressed the 
opinion that the wisps are thermal instabilities in the post-shock outflow from
the pulsar, with the shock itself either unobserved or attributed to
one or another of the time variable features seen in the HST pictures.  In this
case, the observed outflow velocity of the wisp features is the fluid
flow speed. The main flaw
in this view is that within the Kennel and Coroniti model, from which the 
suggestion  derives, the cooling time is too long ($\sim $ 10 years) for 
the particles which mainly contribute
to the pressure, much longer than the weeks to months
needed to explain the variations.
Chedia {\it et al} (1997) suppose the wisps to be drift waves 
in a low energy pair plasma (whose provenance is not otherwise explained).
excited by a $\gamma \sim 10^6$  ion beam from the pulsar - these authors
assume no shock forms, which is contrary to the known dynamics of
a relativistic ion beam in the magnetized plasma. Begelman (1998b) hypothesizes
the wisps to be travelling surface waves excited by the interaction
between an axisymmetric equatorial outflow and a higher latitude outflow
traveling with a different four velocity, a model which is not very
specific about observational consequences that could discriminate it
from other ideas.  Presumably still more suggestions will be forthcoming
as the quality of the observations continues to improve.

\section{Other Plerions}

One may well ask whether other filled image SNR (``plerions'', composite SNR) 
have the same physics.  Unfortunately, none of them have been 
sufficiently well studied to bring the kind of physical modeling described
here to bear.
We saw at this meeting the striking advances in X-ray detections of plerions,
which, when coupled with existing and new radio observations, allow us to
begin asking simple physical questions, such as whether spectral ``breaks''
between radio and X-ray observations tell us anything about the age
of the system, assuming a single mechanism injects a single power law
distribution at all energies into the plerion.  But until the optical and IR
spectral regions are filled in, until detections are made at energies high 
enough to allow one to follow possible variability of the acceleration 
mechanism, and until high angular resolution imaging allows one to study 
the variable features associated with the acceleration physics, one will be
left with only the wonderful example of the Crab as the test of physical 
theories of pulsars as particle accelerators.  The chance that this system's
physics does not typify all the pulsar-plerion pairs known or to be found, and
by extension might not typify the excitation of nonthermal activity by other
central compact objects (such as the jets driven by black holes in AGN),
underlines the need to advance observations of a larger number of 
plerions across the whole spectrum (especially of the younger,
calorimetric systems), with the highest possible 
angular imaging and with
sufficient temporal coverage to follow the variations of the fine scale
structure which are so revealing of the compact object-surrounding world
interaction marvelously exhibited by the Crab. It {\it does} behoove us theorists
to turn some of the insights laboriously gleaned from the Crab into predictions
for other plerions, especially the young ones with calorimetric
morphology. I'm sure there will be plenty of surprises for all of us.

\acknowledgements

I am particularly indebted to Jeff Hester for energetic discussions which
clarified my understanding of the flow geometry around the Crab Pulsar,
as well as many other entertaining aspects of astrophysics and of life, and to 
Anatoly Spitkovsky for illuminating collaboration and discussion. The 
research described here has been supported by NSF grant AST 9528271 and by
NASA grant NAG 5-3073, and in part by the generosity of California's taxpayers.

\end{document}